\documentclass[aps,prb,twocolumn,showpacs,superscriptaddress,groupedaddress]{revtex4}  
\usepackage{graphicx}  
\usepackage{graphics}
\usepackage{amssymb}   
\usepackage{amsmath}
\usepackage{amsfonts}
\usepackage{amstext}
\usepackage{color}

\newcommand{\ns}[0]{n_{\mathrm{s}}} 
\newcommand{\Ah}[0]{A_{\mathrm{h}}} 
\newcommand{\SVh}[0]{S_{V,\mathrm{h}}} 
\newcommand{\SVhc}[0]{S_{V,\mathrm{hc}}} 
\newcommand{\SVw}[0]{S_{V,\mathrm{w}}} 
\newcommand{\SVwexp}[0]{S_{V,\mathrm{w}}^\mathrm{exp}} 
\newcommand{\SVwsim}[0]{S_{V,\mathrm{w}}^\mathrm{sim}} 
\newcommand{\SVm}[0]{S_{V,\mathrm{m}}} 
\newcommand{\SVohm}[0]{S_{V,\Omega}} 
\newcommand{\SVamp}[0]{S_{V,\mathrm{amp}}} 
\newcommand{\SV}[0]{S_{V}}
\newcommand{\kB}[0]{k_\mathrm{B}}
\newcommand{\Ramp}[0]{R_\mathrm{amp}}
\newcommand{\Rh}[0]{R_\mathrm{h}} 
\newcommand{\Rheff}[0]{R_\mathrm{h,eff}} 
\newcommand{\Rohm}[0]{R_{\Omega}} 
\newcommand{\Rohmeff}[0]{R_{\Omega,\mathrm{eff}}} 
\newcommand{\Rin}[0]{R_\mathrm{in}} 
\newcommand{\Tamp}[0]{T_\mathrm{amp}}
\newcommand{\Tbath}[0]{T_\mathrm{bath}}
\newcommand{\Tlat}[0]{T_\mathrm{lat}}
\newcommand{\Te}[0]{T_\mathrm{e}}
\newcommand{\Pe}[0]{P_\mathrm{e}}
\newcommand{\Ih}[0]{I_\mathrm{h}}
\newcommand{\noise}[0]{\cdot 10^{-18}~\mathrm{V}^2/\mathrm{Hz}} 

\begin{document}

\title{Noise thermometry in narrow 2D electron gas heat baths connected to a quasi-1D interferometer}
\date{\today}

\author{Sven S.\ Buchholz}\email[electronic address: ]{sven.buchholz@physik.hu-berlin.de}
\affiliation{Neue Materialien, Institut f\"ur Physik, Humboldt-Universit\"at zu Berlin, 12489 Berlin, Germany}
\affiliation{Werkstoffe und Nanoelektronik, Ruhr-Universit\"at Bochum, 44780 Bochum, Germany}
\author{Elmar Sternemann}
\affiliation{Experimentelle Physik 2, Technische Universit\"at Dortmund, 44227 Dortmund, Germany}
\affiliation{Werkstoffe und Nanoelektronik, Ruhr-Universit\"at Bochum, 44780 Bochum, Germany}
\author{Olivio Chiatti}
\affiliation{Neue Materialien, Institut f\"ur Physik, Humboldt-Universit\"at zu Berlin, 12489 Berlin, Germany}
\author{Dirk Reuter}
\affiliation{Angewandte Festk\"orperphysik, Ruhr-Universit\"at Bochum, 44780 Bochum, Germany}
\author{Andreas D.\ Wieck}
\affiliation{Angewandte Festk\"orperphysik, Ruhr-Universit\"at Bochum, 44780 Bochum, Germany}
\author{Saskia F.\ Fischer}\email[electronic address: ]{saskia.fischer@physik.hu-berlin.de}
\affiliation{Neue Materialien, Institut f\"ur Physik, Humboldt-Universit\"at zu Berlin, 12489 Berlin, Germany}
\affiliation{Werkstoffe und Nanoelektronik, Ruhr-Universit\"at Bochum, 44780 Bochum, Germany}

\begin{abstract}
Thermal voltage noise measurements are performed in order to determine the electron temperature in nanopatterned channels of a GaAs/AlGaAs heterostructure at bath temperatures of 4.2 and 1.4~K. Two narrow two-dimensional (2D) heating channels are connected by a quasi-1D quantum interferometer. Under dc current heating of the electrons in one heating channel, we perform cross-correlated noise measurements locally in the directly heated channel and nonlocally in the other channel, which is indirectly heated by hot electron diffusion across the quasi-1D connection. The temperature dependence of the electron energy-loss rate is reduced compared to wider 2D systems. Under nonlocal current heating, which establishes a thermal gradient across the quantum interferometer, we show the decoherence in this structure by Aharonov-Bohm measurements.
\end{abstract}

\maketitle 

\section{Introduction}\label{sec:introduction}
\vspace*{-0.3cm}

In recent years, research activities in the field of nanostructured materials have been increasingly focused on thermoelectric properties and thermal non-equilibrium.\cite{niel11} In particular, the creation and detection of thermal gradients and the determination of lattice and charge carrier temperatures at the nanoscale remain crucial issues.
Semiconductor heterostructures may be prepared as model systems for thermoelectric investigations representing two-dimensional (2D), 1D, and 0D charge carrier systems. Electrical thermometry methods have been implemented on the basis of resistance and mobility measurements,\cite{wenn86,stor90,mitt96} Shubnikov-de Haas (SdH) measurements,\cite{hira86,mani87,ma91,ridl91,appl98} quantum point contacts (QPCs),\cite{mole90a,appl98,chia06,pros07} and quantum dots (QDs)\cite{star93,sche05,hoff09} for such low-dimensional systems. However, most of these methods are applicable only at temperatures below roughly 20~K. Thermometry via mobility measurements is limited due to the contribution of impurity scattering,\cite{mani87,pros07} and SdH measurements require a magnetic field which alters the density of states and possibly the energy relaxation.\cite{appl98} SdH, QPC and QD thermometry are limited to low lattice temperatures due to thermal smearing of the discrete energy states.

Applying the above thermometry methods enables to study the charge carrier energy-loss to the lattice, which provides fundamental information about electron-phonon interactions. Whereas much effort has been made in the field of 2D charge carrier systems, mostly in the form of wide Hall bar structures,\cite{ridl91,appl98,pros07} only a few experiments focus on the transition to 1D systems where electron-phonon interactions may be altered.\cite{kurd95,suga02,pras04}

Here, we apply electronic noise measurements as a direct method for the determination of the charge carrier temperature in a particularly narrow 2D GaAs/AlGaAs structure. Thermal (Johnson-Nyquist) noise measurements are applicable to different materials with a wide range of operating temperatures, such as diffusive metal films and wires,\cite{rouk85,henn97} and semiconductors hosting high mobility 3D,\cite{eckh03} 2D,\cite{kurd95} and (quasi)-1D electronic systems.\cite{kurd95,kuma96}

We fabricated a device consisting of a quasi-1D quantum interferometer integrated between two narrow 2D heating channels in order to create a temperature difference between the electron reservoirs (heat baths). In these reservoirs, we measured thermal noise locally and nonlocally. The interferometer allows us to study the influence of nonlocal heating on the coherence of electrons in the quantum structure.

The electron system is heated to a temperature above the lattice temperature by means of the current heating technique.\cite{syme89,gall90,mole90a} We extract the electron temperature in the narrow 2D channels for different heating currents, and we find a reduced temperature dependence of the electron energy-loss rate compared to wide 2D electron gases (2DEGs). Resistor network simulations of the thermal noise allow us to determine noise contributions of individual parts of the sample and of the external circuitry. Additionally, by Aharonov-Bohm (AB) measurements, we show a decoherence effect in the interferometer on the basis of nonlocal current heating.

\vspace*{-0.3cm}
\section{Experimental Details}\label{sec:experiment}
\vspace*{-0.3cm}

\begin{figure}[t]
\begin{center}
\includegraphics[]{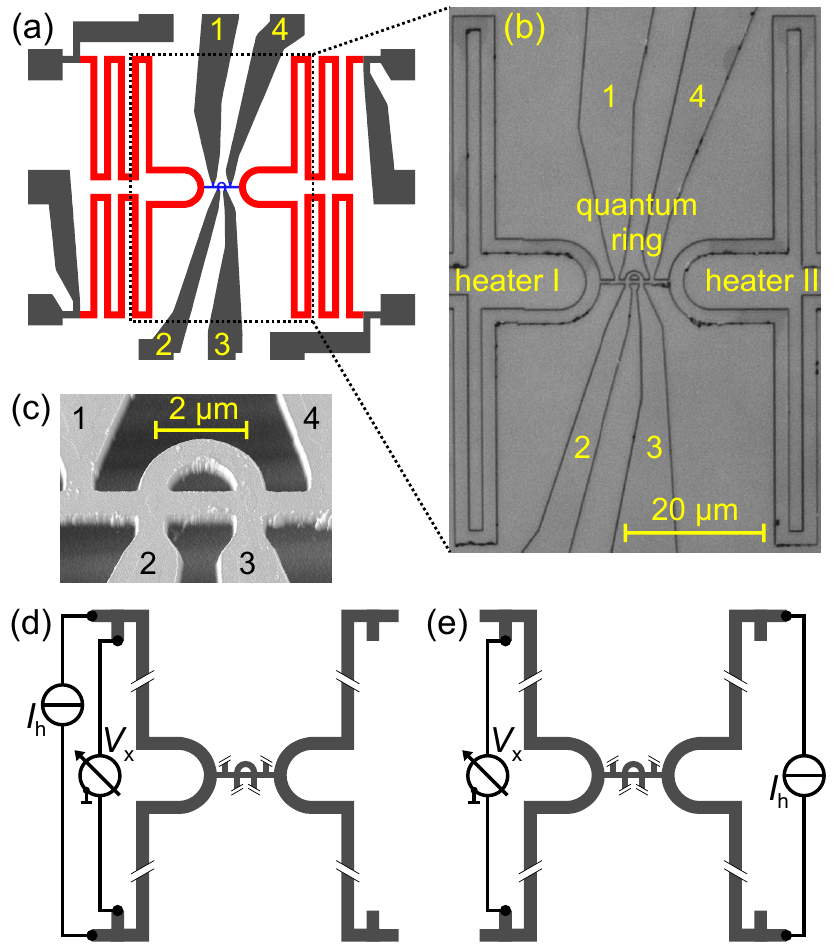}
\caption{(Color online) 
(a) Sample layout (to scale) with four-terminal heating channels as meander structures to the left and the right of the quantum ring. (b) Microscopic photograph of the sample as indicated in (a). (c) Atomic force microscope image of the quantum ring. (d,e) Schemes of the measurement setups (not to scale) for (d) local and (e) nonlocal heating via $\Ih$ and noise measurement $V_\mathrm{x}$.
}
\label{fig1}
\end{center}
\end{figure}

A schematic of the device is depicted in Fig.~\ref{fig1}(a). The two narrow 2D heating channels are nominally identical and connected by a quasi-1D quantum ring. The device was prepared from a GaAs/AlGaAs heterostructure with a 2DEG 110 nm below the surface, using electron beam lithography and 85-nm-deep wet-chemical etching. The 2D electron density and mobility are $\ns=2.07 \cdot 10^{11}$~cm$^{-2}$ and $\mu=2.43\cdot10^6$~cm$^2/\mathrm{Vs}$ at $T=4.2$~K in the dark. The heating channels - labeled 'heater I' and 'heater II' in Fig.~\ref{fig1}(b) - are geometrically $2~\mu$m wide and $410~\mu$m long. This ensures a high thermal noise signal over the background of the total parasitic noise. SdH measurements along the heating channels yield an electron density of $\ns=1.84 \cdot 10^{11}$~cm$^{-2}$, where we attribute the deviation from the above-mentioned sheet density to the lateral confinement of 2~$\mu$m. A non-alloyed gold flake (not shown) on heating channel II remained from the lift-off process but does not influence the electron transport properties.

The quantum wires defining the ring (Fig.~\ref{fig1}(c)) and its leads to the heating channels are geometrically 570~nm wide. From separate measurements on simple quantum wires (widths 350 to 550~nm), as well as four-terminal resistance measurements along the quantum ring, we estimate that about 10 modes of the quasi-1D subband structure are populated in the quantum ring in equilibrium.


Noise measurements were performed in a $^4$He cryostat at bath temperatures of $\Tbath\geq 1.4$~K and recorded with an Agilent 89410A spectrum analyzer. At $T=1.4$~K, the heating channels have a four-terminal resistance of $\Rh\approx 6$~k$\Omega$, which corresponds to a Nyquist noise of less than $10^{-18}$~V$^2/\mathrm{Hz}$. In order to increase the noise signal of the heating channels above the noise of the spectrum analyzer ($\approx 10^{-16}$~V$^2/\mathrm{Hz}$), we used two low-noise voltage preamplifiers with a voltage gain of $10^3$ (Signal Recovery 5184). Cross-correlated measurements were applied to reduce noise contributions from the preamplifiers.\cite{samp99} 

We measured the noise spectrum along heater I in two different heating current setups allowing for local (Fig.~\ref{fig1}(d)) and nonlocal (Fig.~\ref{fig1}(e)) heating. In the local setup, we measured the thermal noise of heater I to which the heating current was applied. In the nonlocal setup, thermal noise was measured in heater I, while the heating current was driven through heater II on the other side of the quantum ring. The heating current was applied via a battery-driven voltage source with a 1~M$\Omega$ series resistor on each side of the source. A 1~$\mu$F capacitor to ground was attached on each side in order to reduce parasitic coupling effects. 

We chose the resistance of the heating channel such that its noise contribution exceeds that of the remaining measurement circuitry. The length of the heating channels exceeds the electron phase coherence and energy relaxation lengths, which yields a diffusive transport regime, where the electron temperature $\Te$ can be deduced from the thermal white noise $\SVw$ by means of the Nyquist formula. If heated by a current, the electrons are no longer in equilibrium with the lattice and share the heat energy among themselves through electron-electron interactions. Energy relaxation takes place via phonon emission and the diffusion to cold reservoirs.

In order to estimate individual thermal noise contributions, we simulated the sample and the circuitry in a SPICE model (Cadence PSpice). Next to the individual resistive and capacitive parts of the wiring and circuitry, the sample was segmented into discrete resistors accounting for the individual Ohmic parts of the sample, which were determined experimentally by lock-in measurements at the different bath temperatures. 


Noise spectra were recorded for bath temperatures in the range of $\Tbath=1.4$ - 10~K and different local and nonlocal heating currents for a frequency range of 1~Hz to 20~kHz. A recorded voltage noise spectrum $\SV (f)$ results from an average of typically 700 sets of data.

In order to obtain the thermal noise of a heating channel $\SVh$ or the equivalent electron temperature $T_e$ from the total noise spectrum, we analyzed the data as follows:
(a) In order to take parasitic capacities $C_\mathrm{par}$ into account, the measured noise $\SVm (f)$ was corrected by a first-order low-pass $\SV (f)=\SVm (f)(1+(2\pi R C_\mathrm{par})^2)$. We determined $C_\mathrm{par}=800$~nF and 565~nF from experiments with and without the heating circuit attached, respectively. 
(b) The total white noise $\SVw$ was extracted as the average value of $\SV (f)$ in a frequency range, where no $1/f$-noise was visible, i.e. from $f=15$~kHz to 19.25~kHz.
(c) We subtracted noise contributions of Ohmic contacts and the wide 2DEG leads $\SVohm$, the heating circuitry $\SVhc$, and the preamplifiers $\SVamp$ by comparison of $\SVw(\Ih\neq0)$ with $\SVw(\Ih=0)$. $\SVamp$ results from the finite current noise of the preamplifiers. 

All parasitic contributions $S_0$ add to the thermal noise of the heater $\SVh$:
\begin{align*}
\SVw &= \SVh + S_0 = \SVh+\SVohm+\SVhc+\SVamp \\
&=4\kB\Te\Rheff+4\kB\Tbath\Rohmeff+\SVhc\\
&\text{ }\text{ }\text{ }+4\kB\Tamp2\Ramp^{-1}\Rin^2,
\tag{1}\label{eq1}
\end{align*}
where the latter term is valid for $\Rin<<\Ramp$. $\Rin$ is the total input resistance, and $\Rheff$ and $\Rohmeff$ are the effective heater and lead (ohmic contacts and wide 2DEG regions) resistances as seen by the preamplifiers. The SPICE network analysis reveals that we can assume $\Rheff\approx\Rh$, $\Rohmeff\approx\Rohm$ and $\Rin\approx\Rh+\Rohm$ with a maximum error of less than $2~\%$. The preamplifiers with an input resistance of $\Ramp=5$~M$\Omega$ operate at $\Tamp=300$~K, and the factor 2 accounts for the two preamplifiers. At $\Tbath=4.2$~K we determined $\Rh=6.2$~k$\Omega$ and $\Rohm=1.4$~k$\Omega$.

The noise contribution of the heating circuit in the local setup was determined by the simulation as $\SVhc^\mathrm{sim}=0.78\noise$, which is in good agreement with the experimental observations of $\SVhc^\mathrm{exp}=0.73\noise$ and $\SVhc^\mathrm{exp}=0.81\noise$ at $\Tbath=4.2$ and 1.4~K, respectively (see Fig.~\ref{fig2}). With the above resistance values and Eq.~\ref{eq1}, we expect a total white noise at 4.2~K with the heating circuit attached in the local setup but $\Ih=0$ of $\SVw=2.93\noise$, which agrees well with the result of the full simulation $\SVw=2.91\noise$.

\vspace*{-0.3cm}
\section{Experimental Results and Discussion}
\vspace*{-0.3cm}

\begin{figure}[t]
\begin{center}
\includegraphics[]{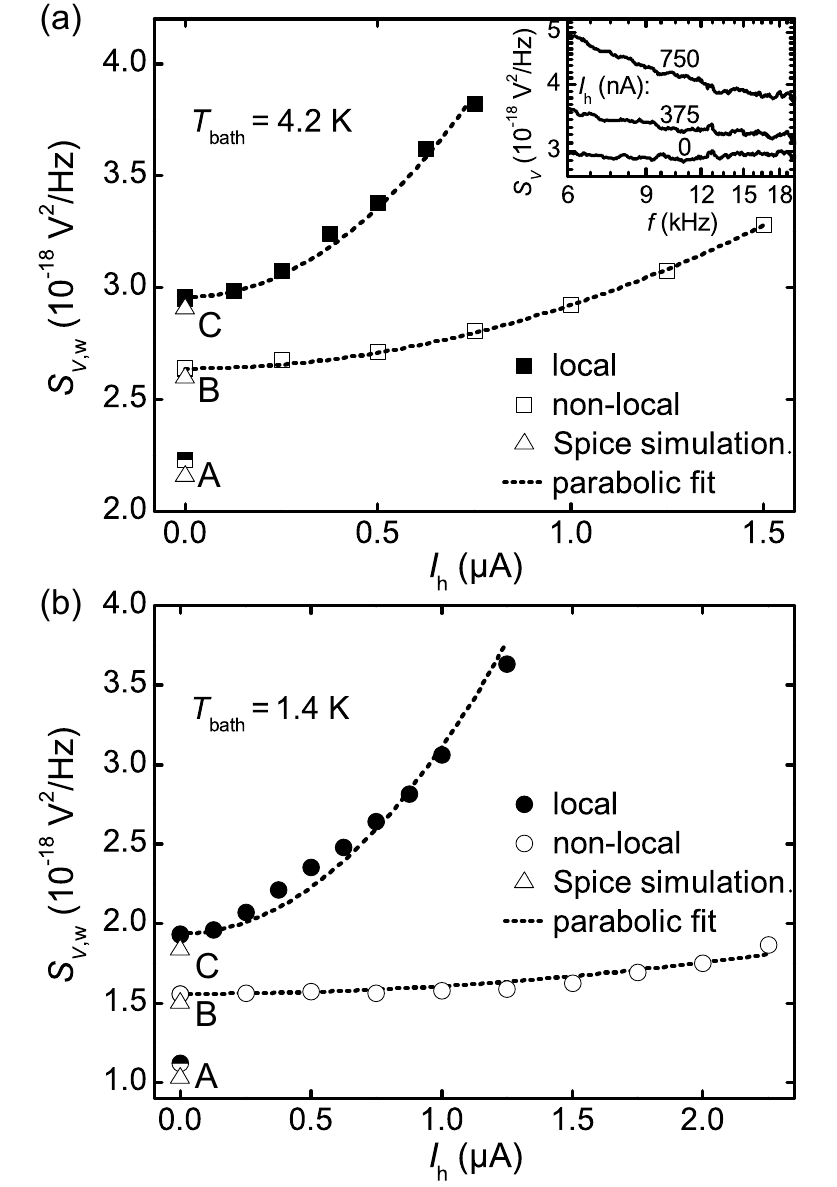}
\caption{
Measurements with parabolic fits and simulations of the voltage noise $\SVw$ for local and nonlocal electron heating. (a) Data at $\Tbath=4.2$~K and (b) $\Tbath=1.4$~K. The filled and empty symbols represent measured values for local and nonlocal heating, respectively. The half-filled symbols (indicated by A) depict measurements without the heating circuit attached. Triangles display results from SPICE simulations, and the broken lines are parabolic fits to the measured data.
}
\label{fig2}
\end{center}
\end{figure}

The determined total white noise $\SVw$ for different heating currents in the local and the nonlocal setup is shown in Fig.~\ref{fig2}(a) for $\Tbath=4.2$~K and in Fig.~\ref{fig2}(b) for $\Tbath=1.4$~K. The size of the symbols represents the measurement accuracy. We ascertained that the heater resistance does not change with the heating current.

The inset of Fig.~\ref{fig2}(a) depicts typical noise spectra $\SV(f)$ for different local heating currents after lowpass correction. For applied heating currents, a $1/f^\gamma$ dependence ($\gamma\leq 1.3$) is visible at $f<13$~kHz. At higher frequencies, white noise dominates. With increasing current, both $1/f$ and white noise components increase. Here, the $1/f$ noise components will not be discussed, but we will consider the white noise component.

Fig.~\ref{fig2} shows the experimentally determined white noise values $\SVwexp$ in the local (filled squares and circles) and nonlocal (empty squares and circles) heating setup, as well as fits to the experimental data (broken lines) and simulated $\SVwsim$ values (triangles). Measured and simulated data without the heating circuit attached are given by symbols labeled A. Values labeled B and C were measured and simulated with the heating circuit attached in the nonlocal and the local setup, respectively. The significant increase of $\SVw$ with the connection of the heating circuit results from the biasing resistors as discussed above.

We will first discuss the data at $\Ih=0$. In order to evaluate the influence of the heating circuit on the total noise quantitatively, we simulated the sample and the measurement circuitry as a resistor network, as explained above, and analyzed the thermal noise. All simulated data points $\SVwsim$ at A, B and C are slightly lower than the corresponding measured values $\SVwexp$, with a maximum deviation of $8\%$. This systematic error corresponds to the thermal noise of a resistance $<10$~$\Omega$ at room temperature and may result from parts of the measurement setup which were not accounted for in the simulation, such as the cables or a higher $\Ramp$ than specified. The good agreement of $\SVwsim$ and $\SVwexp$ shows that the SPICE simulation is a helpful method to investigate the total noise, as well as noise contributions of single components of a complex resistor network.

By applying heating currents, the white noise $\SVw(\Ih\neq0)$ increases with $\Ih$ for both local and nonlocal heating at $\Tbath=4.2$ and 1.4~K. The parabolic best fits in Fig.~\ref{fig2} (broken lines) indicate a quadratic dependence $\Delta\SVw\propto\Ih^2$. In the local heating setup, the electrons of heater I are heated directly. In contrast, under nonlocal heating, the electrons of heater I are heated by hot electron diffusion from heater II across the quantum ring. Phonon mediated contributions to heat transfer were investigated in an additional device (same heterostructure), which possesses a gate electrode to locally deplete the electron system between two heating channels. In this device, a noise increase under nonlocal current heating could only be observed if the electron system was conducting. Once the electron system was depleted, no noise dependence on the heating current was detected. Hence, we expect phonon mediated heat transfer to be negligible in our devices at such low temperatures. 

\begin{figure}[t]
\begin{center}
\includegraphics[]{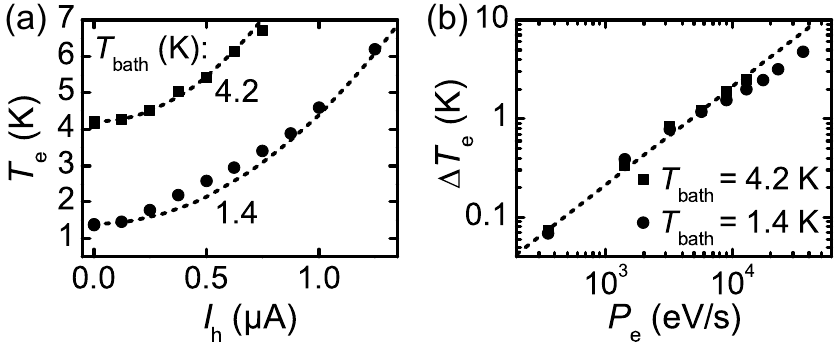}
\caption{
(a) Electron temperature $\Te$ as a function of the heating current $\Ih$ in the local setup for $\Tbath=4.2$ and 1.4~K. The symbols represent data derived from noise measurements, whereas the lines result from parabolic fits. (b) The difference between the electron and the bath temperature $\Delta\Te=\Te-\Tbath$ as a function of the dissipated power per electron $\Pe$, for $\Tbath=4.2$ and 1.4~K (local heating setup). The line results from the parabolic fit for $\Tbath=4.2$~K.
}
\label{fig3}
\end{center}
\end{figure}

Fig.~\ref{fig3}(a) depicts $\Te$ as a function of $\Ih$ for the local heating setup, as determined from Eq.~\ref{eq1}. In addition to the measurement data (symbols), parabolic fits are displayed. In a first approximation, we observe $\Delta\Te\propto\Ih^2$, as expected for Joule heating and observed in other GaAs/AlGaAs 2DEGs.\cite{mole94,maxi04,pros07} For the nonlocal heating setup, we do not attempt an analogous determination of $\Te$ due to the unknown temperature gradient in the narrow channel. 

Commonly, the current is converted to the dissipated power per electron $\Pe=\Ih^2\Rh/(\ns\Ah)$ with $\Ah$ the area of the heating channel. In a steady state, $\Pe$ is taken as the net power transfer from the electrons to the lattice $\Pe=\dot{Q}(\Te)-\dot{Q}(\Tlat)$, \cite{ridl91,kurd95,appl98,pros07} where $\Tlat$ is the lattice temperature and $\dot{Q}$ denotes the energy-loss rate. Fig.~\ref{fig3}(b) shows the electron temperature increase $\Delta\Te=\Te-\Tbath$ as a function of $\Pe$ in a full-logarithmic graph for the local setup. The line gives the parabolic fit for $\Tbath=4.2$~K. While $\Delta\Te$ scales approximately linearly with $\Pe$ for low heating powers, it deviates significantly for $\Pe>7\cdot 10^3$~eV/s, since $\Delta\Te<<\Tlat$ is no longer satisfied, as discussed previously.\cite{pros07} We point out that the temperature increase is almost independent of the bath temperature ($\Tbath=4.2$ and 1.4~K). The absolute electron heating in the chosen $\Pe$-range is in good agreement with data determined via SdH measurements in GaAs/AlGaAs low dimensional electron systems. \cite{hira86,kurd95}
\begin{figure}[t]
\begin{center}
\includegraphics[]{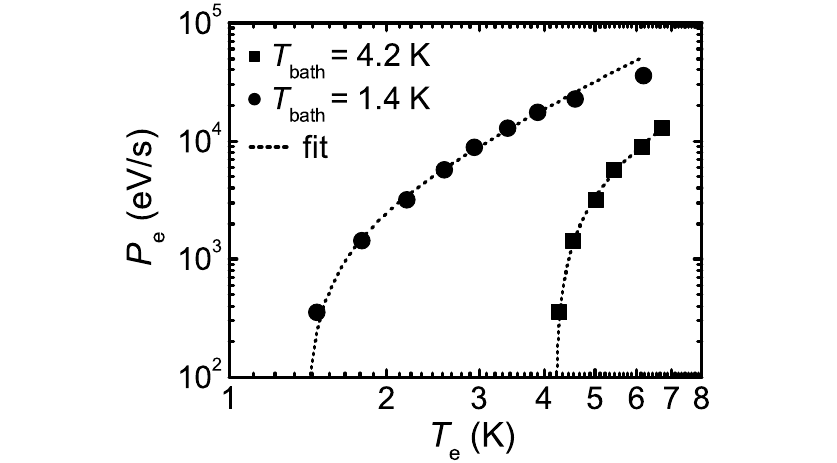}
\caption{
Dissipated power per electron $\Pe$ as a function of the experimentally determined electron temperature $\Te$ in the local heating setup for $\Tbath=4.2$~K and 1.4~K. The broken lines are best fits to $\Pe=A(\Te^{2.2}-\Tbath^{2.2})$.
}
\label{fig4}
\end{center}
\end{figure}

In order to investigate the scattering mechanisms in the heated electron channel, we plot the experimentally determined data for local heating as $\Pe(\Te)$ in Fig.~\ref{fig4}. Here, the broken lines are fits to $\Pe=A(\Te^n-\Tbath^n)$ (two-bath-model\cite{wenn86}) and yield approximately $n=2.2$ for both $\Tbath$, and $A=985$ and $A=300$~eV$/$s$\mathrm{K}^{2.2}$ for $\Tbath=4.2$ and 1.4~K, respectively. For $\Tbath=1.4$~K, the data for $\Te>4$~K deviate from the fit and were not included in the fit. We assume $\Tbath=\Tlat$ since the device is thermally well anchored. From the exponent of $T^n$ in the two-bath-model, information about electron-phonon interactions can be deduced, as discussed in detail elsewhere.\cite{ridl91,appl98,pros07} Whereas investigations of the electron energy-loss rates in GaAs/AlGaAs 2DEGs at $\Tlat=1$~-~10~K by SdH measurements typically yield a $T^2$ or $T^3$-dependence\cite{hira86,ridl91,appl98}, experiments on the basis of QPC thermometry showed a $T^5$-behavior in the same temperature regime, as well as for temperatures in the mK-range.\cite{appl98,pros07} From the latter experiments, it was deduced that the dominant scattering mechanisms in the Gr\"uneisen-Bloch regime are acoustic phonon scattering via a screened piezoelectric potential\cite{appl98} or an unscreened deformation potential.\cite{pros07} $n=3$ suggests scattering via an unscreened piezoelectric potential. $n=2$, if diffusion to cold contacts dominates the energy relaxation.\cite{appl98} $n=1$ is observed in the equipartition regime at higher temperatures of around 20 - 40~K.\cite{ridl91}

The $T^{2.2}$-dependence observed here may suggest that electrons and phonons interact in an intermediate state between the Gr\"uneisen-Bloch and the equipartition regime. However, we point out that the 2~$\mu$m wide heating channels are particularly narrow compared with those investigated previously. Investigations of electron-phonon interactions in 2DEGs were performed in structures of typically several 10~$\mu$m width or wider.\cite{ridl91,appl98,pros07} 

In GaAs based devices, structures of a width less than roughly 1~$\mu$m show signatures of the 1D regime in transport, where samples of width between 50 and 300~nm yield clear quantized conductance at temperatures of a few K.\cite{kris98b,apet02} Structures of several $\mu$m width, on the other hand, are distinct 2D systems whose fundamental parameters are investigated by SdH and quantum Hall measurements. A system of 2~$\mu$m width, as investigated here, reveals its 2D character in simple transport measurements since the subband separation of its 1D energy states is in the range of tens to hundreds of $\mu$eV and thus below the thermal smearing. However, the density of states may be slightly altered toward 1D characteristics, similar to the low-filed SdH effect. One signature for a 2D regime close to the transition to 1D is the decreased electron density in the narrow heating channels compared to wide 2D regions, as described in Sec.~II.

For 1D systems, a reduction of the power law exponent $n$ in the Gr\"uneisen-Bloch regime was predicted\cite{bock90,shik93,das93,kuba07} and experimentally observed in etched InGaAs wires (etching widths between 25~nm and 1~$\mu$m).\cite{suga02,pras04} Under perpendicular magnetic fields, a reduction of $n$ in a 2DEG has been observed, and it was suggested that the momentum transfer of electron-phonon interactions was restricted due to Landau Levels.\cite{appl98} In analogy to this magnetic confinement of the electron-phonon interactions, in 1D systems the electronic confinement with its associated modification of the density of states could restrict electron scattering and thus modify the temperature dependence of electron-phonon interactions.\cite{shik93,das93} In the etched, narrow heating channels investigated here, this effect may play a role. We exclude a significant contribution to the energy relaxation by the wide 2DEG leading to the Ohmic contacts since the heating channel length exceeds the hot electron diffusion length. However, the systematic investigation of the energy-loss rate dependence on the channel width, ranging from wide 2D- to narrow 1D-systems, remains a prospect for future experiments.

For the investigation of the impact of nonlocal current heating on decoherence, we applied a heating current in one heating channel and detected the decoherence by AB measurements in the quantum interferometer. The sample was cooled down in a dilution refrigerator to the base temperature of $\Tbath=20$~mK. Noise measurements could not be performed in this cryostat. While driving the dc heating current $\Ih$ through heater I, we measured the quantum ring's four-terminal resistance $R_{41,32}=V_{32}/I_{41}$ by lock-in technique as a function of the perpendicular magnetic field $B$, exploiting the AB interference effect.\cite{hans01,lin10,buch10}

\begin{figure}[t]
\begin{center}
\includegraphics[]{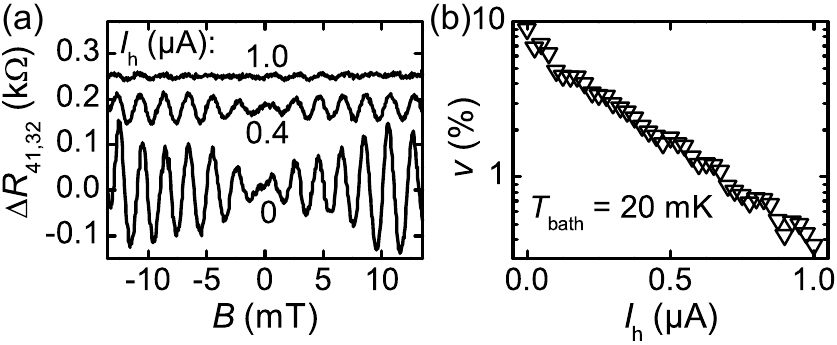}
\caption{
Aharonov-Bohm interference measurements and analysis under current heating through heater I. (a) Oscillatory part of the four-terminal magnetoresistance $R_{41,32}$ for different heating currents $\Ih$, offset 
for clarity. (b) Interference visibility $v$ as a function of $\Ih$. The error amounts to about $\pm 0.3\%$. $R_{41,32}$ was measured in a perpendicular magnetic field $B$ at $\Tbath=20$~mK.
}
\label{fig5}
\end{center}
\end{figure}

Fig.~\ref{fig5}(a) shows the oscillatory part of the magnetoresistance $R_{41,32}$ for three different $\Ih$ after subtraction of the background resistance. For all traces, the resistance oscillates regularly with $B$ with an $h/e$-period, in accordance to the quantum ring geometry. It is noteworthy that the oscillation phase does not change for all applied $\Ih$, i.e. phase rigidity is not lifted by current heating in this setup.\cite{buch10} With increasing $\Ih$, the oscillation amplitude decreases until the AB oscillations are fully suppressed for $\Ih>1.2~\mu$A.

The AB oscillation visibility $v=(R_\mathrm{max}-R_\mathrm{min})/(R_\mathrm{max}+R_\mathrm{min})$ is plotted as a function of $\Ih$ in Fig.~\ref{fig5}(b). The approximately linear decrease of $v$ with $\Ih$ in the semi-logarithmic graph suggests the relation $v(\Ih)=v_0 \exp(-\alpha_\mathrm{h}\Ih)=v_0 \exp(-L/L_\phi)$, with $\alpha_\mathrm{h}$ the fitting parameter, $L$ the mean length of the interferometer arms, and $L_\phi$ the electron phase breaking length.\cite{hans01,buch10}

Under an applied heating current through heater I, hot electrons are created in this reservoir next to the quantum wire interferometer, and a temperature difference between the two heat baths (channel 1 and 2) is invoked (thermal gradient). This leads to the diffusion of hot electrons across the quasi-1D structure, which raises the local electron temperature in the interferometer. Previous works have investigated electron dephasing in quasi-1D AB interferometers by an increase of the lattice temperature and have shown the relation $v(\Tlat)=v_0 \exp(-\alpha_T\Tlat)$, which leads to $L_\phi\propto \Tlat^{-1}$.\cite{hans01,lin10,buch10} Here, we find $L_\phi\propto \Ih^{-1}$, in analogy, and conclude that dephasing with $\Ih$ is induced by the elevated electron temperature in the interferometer via increased electron-electron scattering and thermal averaging.\cite{hans01,lin10,buch10}

\vspace*{-0.3cm}
\section{Summary}
\vspace*{-0.3cm}

In conclusion, we have applied voltage noise thermometry to a particularly narrow 2D GaAs/AlGaAs electron system at bath temperatures of 4.2 and 1.4~K. Electrons were heated by the application of a dc current, and the thermal noise was measured in a cross-correlation setup. The device consists of two heating channels (heat baths) with a quasi-1D quantum interferometer in-between. We performed local noise measurements in the directly heated channel and nonlocal measurements in the indirectly heated channel, and we found the same functional dependence of the thermal noise on the heating current in the local and the nonlocal heating setup. The indirect heating is explained by hot electron diffusion through the quasi-1D interferometer. The temperature dependence of the electron energy-loss rate of $T^{2.2}$ is lower than that observed in previous investigations at 2DEGs. This may result from the confinement of the electron system to a narrow 2DEG with a slightly altered density of states evoking restrictions in phase space of electron-phonon interactions. We demonstrate that an indirect current heating can be successfully employed as a means to establish thermal gradients between heat baths connected to (quasi-) 1D quantum circuits. The effect of indirect current heating on the electron decoherence in the quantum interferometer was investigated by Aharonov-Bohm measurements, and an exponential decay of the visibility with an increasing heating current was observed.

\vspace*{-0.3cm}
\section*{Acknowledgments}
\vspace*{-0.3cm}

The authors gratefully acknowledge financial support from the Deutsche Forschungsgemeinschaft (DFG) within the priority program SPP1386. S.F.F. is grateful for support from the 'Junges Kolleg,' North Rhine-Westphalia Academy for Sciences and Arts. D.R. and A.D.W. acknowledge support from the DFG SPP1285 and the BMBF QuaHL-Rep 01BQ1035. We greatly appreciate the scientific and technical support from Prof. U. Kunze.

\bibliography{Svens_bibliography}

\begin{thebibliography}{36}%
\makeatletter
\providecommand \@ifxundefined [1]{%
 \@ifx{#1\undefined}
}%
\providecommand \@ifnum [1]{%
 \ifnum #1\expandafter \@firstoftwo
 \else \expandafter \@secondoftwo
 \fi
}%
\providecommand \@ifx [1]{%
 \ifx #1\expandafter \@firstoftwo
 \else \expandafter \@secondoftwo
 \fi
}%
\providecommand \natexlab [1]{#1}%
\providecommand \enquote  [1]{``#1''}%
\providecommand \bibnamefont  [1]{#1}%
\providecommand \bibfnamefont [1]{#1}%
\providecommand \citenamefont [1]{#1}%
\providecommand \href@noop [0]{\@secondoftwo}%
\providecommand \href [0]{\begingroup \@sanitize@url \@href}%
\providecommand \@href[1]{\@@startlink{#1}\@@href}%
\providecommand \@@href[1]{\endgroup#1\@@endlink}%
\providecommand \@sanitize@url [0]{\catcode `\\12\catcode `\$12\catcode
  `\&12\catcode `\#12\catcode `\^12\catcode `\_12\catcode `\%12\relax}%
\providecommand \@@startlink[1]{}%
\providecommand \@@endlink[0]{}%
\providecommand \url  [0]{\begingroup\@sanitize@url \@url }%
\providecommand \@url [1]{\endgroup\@href {#1}{\urlprefix }}%
\providecommand \urlprefix  [0]{URL }%
\providecommand \Eprint [0]{\href }%
\providecommand \doibase [0]{http://dx.doi.org/}%
\providecommand \selectlanguage [0]{\@gobble}%
\providecommand \bibinfo  [0]{\@secondoftwo}%
\providecommand \bibfield  [0]{\@secondoftwo}%
\providecommand \translation [1]{[#1]}%
\providecommand \BibitemOpen [0]{}%
\providecommand \bibitemStop [0]{}%
\providecommand \bibitemNoStop [0]{.\EOS\space}%
\providecommand \EOS [0]{\spacefactor3000\relax}%
\providecommand \BibitemShut  [1]{\csname bibitem#1\endcsname}%
\let\auto@bib@innerbib\@empty
\bibitem [{\citenamefont {Nielsch}\ \emph {et~al.}(2011)\citenamefont
  {Nielsch}, \citenamefont {Bachmann}, \citenamefont {Kimling},\ and\
  \citenamefont {B\"ottner}}]{niel11}%
  \BibitemOpen
  \bibfield  {author} {\bibinfo {author} {\bibfnamefont {K.}~\bibnamefont
  {Nielsch}}, \bibinfo {author} {\bibfnamefont {J.}~\bibnamefont {Bachmann}},
  \bibinfo {author} {\bibfnamefont {J.}~\bibnamefont {Kimling}}, \ and\
  \bibinfo {author} {\bibfnamefont {H.}~\bibnamefont {B\"ottner}},\ }\href@noop
  {} {\bibfield  {journal} {\bibinfo  {journal} {Adv. Engery Mater.}\ }\textbf
  {\bibinfo {volume} {1}},\ \bibinfo {pages} {713} (\bibinfo {year}
  {2011})}\BibitemShut {NoStop}%
\bibitem [{\citenamefont {Wennberg}\ \emph {et~al.}(1986)\citenamefont
  {Wennberg}, \citenamefont {Ytterboe}, \citenamefont {Gould}, \citenamefont
  {Bozler}, \citenamefont {Klem},\ and\ \citenamefont {Morkoc}}]{wenn86}%
  \BibitemOpen
  \bibfield  {author} {\bibinfo {author} {\bibfnamefont {A.~K.~M.}\
  \bibnamefont {Wennberg}}, \bibinfo {author} {\bibfnamefont {S.~N.}\
  \bibnamefont {Ytterboe}}, \bibinfo {author} {\bibfnamefont {C.~M.}\
  \bibnamefont {Gould}}, \bibinfo {author} {\bibfnamefont {H.~M.}\ \bibnamefont
  {Bozler}}, \bibinfo {author} {\bibfnamefont {J.}~\bibnamefont {Klem}}, \ and\
  \bibinfo {author} {\bibfnamefont {H.}~\bibnamefont {Morkoc}},\ }\href@noop {}
  {\bibfield  {journal} {\bibinfo  {journal} {Phys. Rev. B}\ }\textbf {\bibinfo
  {volume} {34}},\ \bibinfo {pages} {4409} (\bibinfo {year}
  {1986})}\BibitemShut {NoStop}%
\bibitem [{\citenamefont {Stormer}\ \emph {et~al.}(1990)\citenamefont
  {Stormer}, \citenamefont {Pfeiffer}, \citenamefont {Baldwin},\ and\
  \citenamefont {West}}]{stor90}%
  \BibitemOpen
  \bibfield  {author} {\bibinfo {author} {\bibfnamefont {H.~L.}\ \bibnamefont
  {Stormer}}, \bibinfo {author} {\bibfnamefont {L.~N.}\ \bibnamefont
  {Pfeiffer}}, \bibinfo {author} {\bibfnamefont {K.~W.}\ \bibnamefont
  {Baldwin}}, \ and\ \bibinfo {author} {\bibfnamefont {K.~W.}\ \bibnamefont
  {West}},\ }\href@noop {} {\bibfield  {journal} {\bibinfo  {journal} {Phys.
  Rev. B}\ }\textbf {\bibinfo {volume} {41}},\ \bibinfo {pages} {1278}
  (\bibinfo {year} {1990})}\BibitemShut {NoStop}%
\bibitem [{\citenamefont {Mittal}\ \emph {et~al.}(1996)\citenamefont {Mittal},
  \citenamefont {Wheeler}, \citenamefont {Keller}, \citenamefont {Prober},\
  and\ \citenamefont {Sacks}}]{mitt96}%
  \BibitemOpen
  \bibfield  {author} {\bibinfo {author} {\bibfnamefont {A.}~\bibnamefont
  {Mittal}}, \bibinfo {author} {\bibfnamefont {R.~G.}\ \bibnamefont {Wheeler}},
  \bibinfo {author} {\bibfnamefont {M.~W.}\ \bibnamefont {Keller}}, \bibinfo
  {author} {\bibfnamefont {D.~E.}\ \bibnamefont {Prober}}, \ and\ \bibinfo
  {author} {\bibfnamefont {R.~N.}\ \bibnamefont {Sacks}},\ }\href@noop {}
  {\bibfield  {journal} {\bibinfo  {journal} {Surf. Sci.}\ }\textbf {\bibinfo
  {volume} {361/362}},\ \bibinfo {pages} {537} (\bibinfo {year}
  {1996})}\BibitemShut {NoStop}%
\bibitem [{\citenamefont {Hirakawa}\ and\ \citenamefont
  {Sasaki}(1986)}]{hira86}%
  \BibitemOpen
  \bibfield  {author} {\bibinfo {author} {\bibfnamefont {K.}~\bibnamefont
  {Hirakawa}}\ and\ \bibinfo {author} {\bibfnamefont {H.}~\bibnamefont
  {Sasaki}},\ }\href@noop {} {\bibfield  {journal} {\bibinfo  {journal} {Appl.
  Phys. Lett.}\ }\textbf {\bibinfo {volume} {49}},\ \bibinfo {pages} {889}
  (\bibinfo {year} {1986})}\BibitemShut {NoStop}%
\bibitem [{\citenamefont {Manion}\ \emph {et~al.}(1987)\citenamefont {Manion},
  \citenamefont {Artaki}, \citenamefont {Emanuel}, \citenamefont {Coleman},\
  and\ \citenamefont {Hess}}]{mani87}%
  \BibitemOpen
  \bibfield  {author} {\bibinfo {author} {\bibfnamefont {S.~J.}\ \bibnamefont
  {Manion}}, \bibinfo {author} {\bibfnamefont {M.}~\bibnamefont {Artaki}},
  \bibinfo {author} {\bibfnamefont {M.~A.}\ \bibnamefont {Emanuel}}, \bibinfo
  {author} {\bibfnamefont {J.~J.}\ \bibnamefont {Coleman}}, \ and\ \bibinfo
  {author} {\bibfnamefont {K.}~\bibnamefont {Hess}},\ }\href@noop {} {\bibfield
   {journal} {\bibinfo  {journal} {Phys. Rev. B}\ }\textbf {\bibinfo {volume}
  {35}},\ \bibinfo {pages} {9203} (\bibinfo {year} {1987})}\BibitemShut
  {NoStop}%
\bibitem [{\citenamefont {Ma}\ \emph {et~al.}(1991)\citenamefont {Ma},
  \citenamefont {Fletcher}, \citenamefont {Zaremba}, \citenamefont {D'Iorio},
  \citenamefont {Foxon},\ and\ \citenamefont {Harris}}]{ma91}%
  \BibitemOpen
  \bibfield  {author} {\bibinfo {author} {\bibfnamefont {Y.}~\bibnamefont
  {Ma}}, \bibinfo {author} {\bibfnamefont {R.}~\bibnamefont {Fletcher}},
  \bibinfo {author} {\bibfnamefont {E.}~\bibnamefont {Zaremba}}, \bibinfo
  {author} {\bibfnamefont {M.}~\bibnamefont {D'Iorio}}, \bibinfo {author}
  {\bibfnamefont {C.~T.}\ \bibnamefont {Foxon}}, \ and\ \bibinfo {author}
  {\bibfnamefont {J.~J.}\ \bibnamefont {Harris}},\ }\href@noop {} {\bibfield
  {journal} {\bibinfo  {journal} {Phys. Rev. B}\ }\textbf {\bibinfo {volume}
  {43}},\ \bibinfo {pages} {9033} (\bibinfo {year} {1991})}\BibitemShut
  {NoStop}%
\bibitem [{\citenamefont {for a review~see: B.~K.~Ridley}(1991)}]{ridl91}%
  \BibitemOpen
  \bibfield  {author} {\bibinfo {author} {\bibnamefont {for a review~see:
  B.~K.~Ridley}},\ }\href@noop {} {\bibfield  {journal} {\bibinfo  {journal}
  {Rep. Prog. Phys.}\ }\textbf {\bibinfo {volume} {54}},\ \bibinfo {pages}
  {169} (\bibinfo {year} {1991})}\BibitemShut {NoStop}%
\bibitem [{\citenamefont {Appleyard}\ \emph {et~al.}(1998)\citenamefont
  {Appleyard}, \citenamefont {Nicholls}, \citenamefont {Simmons}, \citenamefont
  {Tribe},\ and\ \citenamefont {Pepper}}]{appl98}%
  \BibitemOpen
  \bibfield  {author} {\bibinfo {author} {\bibfnamefont {N.~J.}\ \bibnamefont
  {Appleyard}}, \bibinfo {author} {\bibfnamefont {J.~T.}\ \bibnamefont
  {Nicholls}}, \bibinfo {author} {\bibfnamefont {M.~Y.}\ \bibnamefont
  {Simmons}}, \bibinfo {author} {\bibfnamefont {W.~R.}\ \bibnamefont {Tribe}},
  \ and\ \bibinfo {author} {\bibfnamefont {M.}~\bibnamefont {Pepper}},\
  }\href@noop {} {\bibfield  {journal} {\bibinfo  {journal} {Phys. Rev. Lett.}\
  }\textbf {\bibinfo {volume} {81}},\ \bibinfo {pages} {3491} (\bibinfo {year}
  {1998})}\BibitemShut {NoStop}%
\bibitem [{\citenamefont {Molenkamp}\ \emph {et~al.}(1990)\citenamefont
  {Molenkamp}, \citenamefont {van Houten}, \citenamefont {Beenakker},
  \citenamefont {Eppenga},\ and\ \citenamefont {Foxon}}]{mole90a}%
  \BibitemOpen
  \bibfield  {author} {\bibinfo {author} {\bibfnamefont {L.~W.}\ \bibnamefont
  {Molenkamp}}, \bibinfo {author} {\bibfnamefont {H.}~\bibnamefont {van
  Houten}}, \bibinfo {author} {\bibfnamefont {C.~W.~J.}\ \bibnamefont
  {Beenakker}}, \bibinfo {author} {\bibfnamefont {R.}~\bibnamefont {Eppenga}},
  \ and\ \bibinfo {author} {\bibfnamefont {C.~T.}\ \bibnamefont {Foxon}},\
  }\href@noop {} {\bibfield  {journal} {\bibinfo  {journal} {Phys. Rev. Lett.}\
  }\textbf {\bibinfo {volume} {65}},\ \bibinfo {pages} {1052} (\bibinfo {year}
  {1990})}\BibitemShut {NoStop}%
\bibitem [{\citenamefont {Chiatti}\ \emph {et~al.}(2006)\citenamefont
  {Chiatti}, \citenamefont {Nicholls}, \citenamefont {Proskuryakov},
  \citenamefont {Lumpkin}, \citenamefont {Farrer},\ and\ \citenamefont
  {Ritchie}}]{chia06}%
  \BibitemOpen
  \bibfield  {author} {\bibinfo {author} {\bibfnamefont {O.}~\bibnamefont
  {Chiatti}}, \bibinfo {author} {\bibfnamefont {J.~T.}\ \bibnamefont
  {Nicholls}}, \bibinfo {author} {\bibfnamefont {Y.~Y.}\ \bibnamefont
  {Proskuryakov}}, \bibinfo {author} {\bibfnamefont {N.}~\bibnamefont
  {Lumpkin}}, \bibinfo {author} {\bibfnamefont {I.}~\bibnamefont {Farrer}}, \
  and\ \bibinfo {author} {\bibfnamefont {D.~A.}\ \bibnamefont {Ritchie}},\
  }\href@noop {} {\bibfield  {journal} {\bibinfo  {journal} {Phys. Rev. Lett.}\
  }\textbf {\bibinfo {volume} {97}},\ \bibinfo {pages} {056601} (\bibinfo
  {year} {2006})}\BibitemShut {NoStop}%
\bibitem [{\citenamefont {Proskuryakov}\ \emph {et~al.}(2007)\citenamefont
  {Proskuryakov}, \citenamefont {Nicholls}, \citenamefont {Hadji-Ristic},
  \citenamefont {Kristensen},\ and\ \citenamefont {Sorensen}}]{pros07}%
  \BibitemOpen
  \bibfield  {author} {\bibinfo {author} {\bibfnamefont {Y.~Y.}\ \bibnamefont
  {Proskuryakov}}, \bibinfo {author} {\bibfnamefont {J.~T.}\ \bibnamefont
  {Nicholls}}, \bibinfo {author} {\bibfnamefont {D.~I.}\ \bibnamefont
  {Hadji-Ristic}}, \bibinfo {author} {\bibfnamefont {A.}~\bibnamefont
  {Kristensen}}, \ and\ \bibinfo {author} {\bibfnamefont {C.~B.}\ \bibnamefont
  {Sorensen}},\ }\href@noop {} {\bibfield  {journal} {\bibinfo  {journal}
  {Phys. Rev. B}\ }\textbf {\bibinfo {volume} {75}},\ \bibinfo {pages} {045308}
  (\bibinfo {year} {2007})}\BibitemShut {NoStop}%
\bibitem [{\citenamefont {Staring}\ \emph {et~al.}(1993)\citenamefont
  {Staring}, \citenamefont {Molenkamp}, \citenamefont {Alphenaar},
  \citenamefont {van Houten}, \citenamefont {Buyk}, \citenamefont {Mabesoone},
  \citenamefont {Beenakker},\ and\ \citenamefont {Foxon}}]{star93}%
  \BibitemOpen
  \bibfield  {author} {\bibinfo {author} {\bibfnamefont {A.~A.~M.}\
  \bibnamefont {Staring}}, \bibinfo {author} {\bibfnamefont {L.~W.}\
  \bibnamefont {Molenkamp}}, \bibinfo {author} {\bibfnamefont {B.~W.}\
  \bibnamefont {Alphenaar}}, \bibinfo {author} {\bibfnamefont {H.}~\bibnamefont
  {van Houten}}, \bibinfo {author} {\bibfnamefont {O.~J.~A.}\ \bibnamefont
  {Buyk}}, \bibinfo {author} {\bibfnamefont {M.~A.~A.}\ \bibnamefont
  {Mabesoone}}, \bibinfo {author} {\bibfnamefont {C.~W.~J.}\ \bibnamefont
  {Beenakker}}, \ and\ \bibinfo {author} {\bibfnamefont {C.~T.}\ \bibnamefont
  {Foxon}},\ }\href@noop {} {\bibfield  {journal} {\bibinfo  {journal}
  {Europhys. Lett.}\ }\textbf {\bibinfo {volume} {22}},\ \bibinfo {pages} {57}
  (\bibinfo {year} {1993})}\BibitemShut {NoStop}%
\bibitem [{\citenamefont {Scheibner}\ \emph {et~al.}(2005)\citenamefont
  {Scheibner}, \citenamefont {Buhmann}, \citenamefont {Reuter}, \citenamefont
  {Kiselev},\ and\ \citenamefont {Molenkamp}}]{sche05}%
  \BibitemOpen
  \bibfield  {author} {\bibinfo {author} {\bibfnamefont {R.}~\bibnamefont
  {Scheibner}}, \bibinfo {author} {\bibfnamefont {H.}~\bibnamefont {Buhmann}},
  \bibinfo {author} {\bibfnamefont {D.}~\bibnamefont {Reuter}}, \bibinfo
  {author} {\bibfnamefont {M.~N.}\ \bibnamefont {Kiselev}}, \ and\ \bibinfo
  {author} {\bibfnamefont {L.~W.}\ \bibnamefont {Molenkamp}},\ }\href@noop {}
  {\bibfield  {journal} {\bibinfo  {journal} {Phys. Rev. Lett.}\ }\textbf
  {\bibinfo {volume} {95}},\ \bibinfo {pages} {176602} (\bibinfo {year}
  {2005})}\BibitemShut {NoStop}%
\bibitem [{\citenamefont {Hoffmann}\ \emph {et~al.}(2009)\citenamefont
  {Hoffmann}, \citenamefont {Nilsson}, \citenamefont {Matthews}, \citenamefont
  {Nakpathomkun}, \citenamefont {Persson}, \citenamefont {Samuelson},\ and\
  \citenamefont {Linke}}]{hoff09}%
  \BibitemOpen
  \bibfield  {author} {\bibinfo {author} {\bibfnamefont {E.~A.}\ \bibnamefont
  {Hoffmann}}, \bibinfo {author} {\bibfnamefont {H.~A.}\ \bibnamefont
  {Nilsson}}, \bibinfo {author} {\bibfnamefont {J.~E.}\ \bibnamefont
  {Matthews}}, \bibinfo {author} {\bibfnamefont {N.}~\bibnamefont
  {Nakpathomkun}}, \bibinfo {author} {\bibfnamefont {A.~I.}\ \bibnamefont
  {Persson}}, \bibinfo {author} {\bibfnamefont {L.}~\bibnamefont {Samuelson}},
  \ and\ \bibinfo {author} {\bibfnamefont {H.}~\bibnamefont {Linke}},\
  }\href@noop {} {\bibfield  {journal} {\bibinfo  {journal} {Nano Lett.}\
  }\textbf {\bibinfo {volume} {9}},\ \bibinfo {pages} {779} (\bibinfo {year}
  {2009})}\BibitemShut {NoStop}%
\bibitem [{\citenamefont {Kurdak}\ \emph {et~al.}(1995)\citenamefont {Kurdak},
  \citenamefont {Tsui}, \citenamefont {Parihar}, \citenamefont {Lyon},\ and\
  \citenamefont {Shayegan}}]{kurd95}%
  \BibitemOpen
  \bibfield  {author} {\bibinfo {author} {\bibfnamefont {C.}~\bibnamefont
  {Kurdak}}, \bibinfo {author} {\bibfnamefont {D.~C.}\ \bibnamefont {Tsui}},
  \bibinfo {author} {\bibfnamefont {S.}~\bibnamefont {Parihar}}, \bibinfo
  {author} {\bibfnamefont {S.~A.}\ \bibnamefont {Lyon}}, \ and\ \bibinfo
  {author} {\bibfnamefont {M.}~\bibnamefont {Shayegan}},\ }\href@noop {}
  {\bibfield  {journal} {\bibinfo  {journal} {Appl. Phys. Lett.}\ }\textbf
  {\bibinfo {volume} {67}},\ \bibinfo {pages} {386} (\bibinfo {year}
  {1995})}\BibitemShut {NoStop}%
\bibitem [{\citenamefont {Sugaya}\ \emph {et~al.}(2002)\citenamefont {Sugaya},
  \citenamefont {Bird}, \citenamefont {Ferry}, \citenamefont {Sergeev},
  \citenamefont {Mitin}, \citenamefont {Jang}, \citenamefont {Ogura},\ and\
  \citenamefont {Sugiyama}}]{suga02}%
  \BibitemOpen
  \bibfield  {author} {\bibinfo {author} {\bibfnamefont {T.}~\bibnamefont
  {Sugaya}}, \bibinfo {author} {\bibfnamefont {J.~P.}\ \bibnamefont {Bird}},
  \bibinfo {author} {\bibfnamefont {D.~K.}\ \bibnamefont {Ferry}}, \bibinfo
  {author} {\bibfnamefont {A.}~\bibnamefont {Sergeev}}, \bibinfo {author}
  {\bibfnamefont {V.}~\bibnamefont {Mitin}}, \bibinfo {author} {\bibfnamefont
  {K.-Y.}\ \bibnamefont {Jang}}, \bibinfo {author} {\bibfnamefont
  {M.}~\bibnamefont {Ogura}}, \ and\ \bibinfo {author} {\bibfnamefont
  {Y.}~\bibnamefont {Sugiyama}},\ }\href@noop {} {\bibfield  {journal}
  {\bibinfo  {journal} {Appl. Phys. Lett.}\ }\textbf {\bibinfo {volume} {81}},\
  \bibinfo {pages} {727} (\bibinfo {year} {2002})}\BibitemShut {NoStop}%
\bibitem [{\citenamefont {Prasad}, \citenamefont {Ferry},\ and\ \citenamefont
  {Wieder}(2004)}]{pras04}%
  \BibitemOpen
  \bibfield  {author} {\bibinfo {author} {\bibfnamefont {C.}~\bibnamefont
  {Prasad}}, \bibinfo {author} {\bibfnamefont {D.~K.}\ \bibnamefont {Ferry}}, \
  and\ \bibinfo {author} {\bibfnamefont {H.~H.}\ \bibnamefont {Wieder}},\
  }\href@noop {} {\bibfield  {journal} {\bibinfo  {journal} {Semicond. Sci.
  Technol.}\ }\textbf {\bibinfo {volume} {19}},\ \bibinfo {pages} {S60}
  (\bibinfo {year} {2004})}\BibitemShut {NoStop}%
\bibitem [{\citenamefont {Roukes}\ \emph {et~al.}(1985)\citenamefont {Roukes},
  \citenamefont {Freeman}, \citenamefont {Germain}, \citenamefont
  {Richardson},\ and\ \citenamefont {Ketchen}}]{rouk85}%
  \BibitemOpen
  \bibfield  {author} {\bibinfo {author} {\bibfnamefont {M.~L.}\ \bibnamefont
  {Roukes}}, \bibinfo {author} {\bibfnamefont {M.~R.}\ \bibnamefont {Freeman}},
  \bibinfo {author} {\bibfnamefont {R.~S.}\ \bibnamefont {Germain}}, \bibinfo
  {author} {\bibfnamefont {R.~C.}\ \bibnamefont {Richardson}}, \ and\ \bibinfo
  {author} {\bibfnamefont {M.~B.}\ \bibnamefont {Ketchen}},\ }\href@noop {}
  {\bibfield  {journal} {\bibinfo  {journal} {Phys. Rev. Lett.}\ }\textbf
  {\bibinfo {volume} {55}},\ \bibinfo {pages} {422} (\bibinfo {year}
  {1985})}\BibitemShut {NoStop}%
\bibitem [{\citenamefont {Henny}\ \emph {et~al.}(1997)\citenamefont {Henny},
  \citenamefont {Birk}, \citenamefont {Huber}, \citenamefont {Strunk},
  \citenamefont {Bachtold}, \citenamefont {Kr\"uger},\ and\ \citenamefont
  {Sch\"onenberger}}]{henn97}%
  \BibitemOpen
  \bibfield  {author} {\bibinfo {author} {\bibfnamefont {M.}~\bibnamefont
  {Henny}}, \bibinfo {author} {\bibfnamefont {H.}~\bibnamefont {Birk}},
  \bibinfo {author} {\bibfnamefont {R.}~\bibnamefont {Huber}}, \bibinfo
  {author} {\bibfnamefont {C.}~\bibnamefont {Strunk}}, \bibinfo {author}
  {\bibfnamefont {A.}~\bibnamefont {Bachtold}}, \bibinfo {author}
  {\bibfnamefont {M.}~\bibnamefont {Kr\"uger}}, \ and\ \bibinfo {author}
  {\bibfnamefont {C.}~\bibnamefont {Sch\"onenberger}},\ }\href@noop {}
  {\bibfield  {journal} {\bibinfo  {journal} {Appl. Phys. Lett.}\ }\textbf
  {\bibinfo {volume} {71}},\ \bibinfo {pages} {773} (\bibinfo {year}
  {1997})}\BibitemShut {NoStop}%
\bibitem [{\citenamefont {Eckhause}\ \emph {et~al.}(2003)\citenamefont
  {Eckhause}, \citenamefont {S\"ulzer}, \citenamefont {Kurdak}, \citenamefont
  {Yun},\ and\ \citenamefont {Morkoc}}]{eckh03}%
  \BibitemOpen
  \bibfield  {author} {\bibinfo {author} {\bibfnamefont {T.~A.}\ \bibnamefont
  {Eckhause}}, \bibinfo {author} {\bibfnamefont {O.}~\bibnamefont {S\"ulzer}},
  \bibinfo {author} {\bibfnamefont {C.}~\bibnamefont {Kurdak}}, \bibinfo
  {author} {\bibfnamefont {F.}~\bibnamefont {Yun}}, \ and\ \bibinfo {author}
  {\bibfnamefont {H.}~\bibnamefont {Morkoc}},\ }\href@noop {} {\bibfield
  {journal} {\bibinfo  {journal} {Appl. Phys. Lett.}\ }\textbf {\bibinfo
  {volume} {82}},\ \bibinfo {pages} {3035} (\bibinfo {year}
  {2003})}\BibitemShut {NoStop}%
\bibitem [{\citenamefont {Kumar}\ \emph {et~al.}(1996)\citenamefont {Kumar},
  \citenamefont {Saminadayar}, \citenamefont {Glattli}, \citenamefont {Jin},\
  and\ \citenamefont {Etienne}}]{kuma96}%
  \BibitemOpen
  \bibfield  {author} {\bibinfo {author} {\bibfnamefont {A.}~\bibnamefont
  {Kumar}}, \bibinfo {author} {\bibfnamefont {L.}~\bibnamefont {Saminadayar}},
  \bibinfo {author} {\bibfnamefont {D.~C.}\ \bibnamefont {Glattli}}, \bibinfo
  {author} {\bibfnamefont {Y.}~\bibnamefont {Jin}}, \ and\ \bibinfo {author}
  {\bibfnamefont {B.}~\bibnamefont {Etienne}},\ }\href@noop {} {\bibfield
  {journal} {\bibinfo  {journal} {Phys. Rev. Lett.}\ }\textbf {\bibinfo
  {volume} {76}},\ \bibinfo {pages} {2778} (\bibinfo {year}
  {1996})}\BibitemShut {NoStop}%
\bibitem [{\citenamefont {Syme}, \citenamefont {Kelly},\ and\ \citenamefont
  {Pepper}(1989)}]{syme89}%
  \BibitemOpen
  \bibfield  {author} {\bibinfo {author} {\bibfnamefont {R.~T.}\ \bibnamefont
  {Syme}}, \bibinfo {author} {\bibfnamefont {M.~J.}\ \bibnamefont {Kelly}}, \
  and\ \bibinfo {author} {\bibfnamefont {M.}~\bibnamefont {Pepper}},\
  }\href@noop {} {\bibfield  {journal} {\bibinfo  {journal} {J. Phys.: Condens.
  Matter}\ }\textbf {\bibinfo {volume} {1}},\ \bibinfo {pages} {3375} (\bibinfo
  {year} {1989})}\BibitemShut {NoStop}%
\bibitem [{\citenamefont {Gallagher}\ \emph {et~al.}(1990)\citenamefont
  {Gallagher}, \citenamefont {Galloway}, \citenamefont {Beton}, \citenamefont
  {Oxley}, \citenamefont {Beaumont}, \citenamefont {Thoms},\ and\ \citenamefont
  {Wilkinson}}]{gall90}%
  \BibitemOpen
  \bibfield  {author} {\bibinfo {author} {\bibfnamefont {B.~L.}\ \bibnamefont
  {Gallagher}}, \bibinfo {author} {\bibfnamefont {T.}~\bibnamefont {Galloway}},
  \bibinfo {author} {\bibfnamefont {P.}~\bibnamefont {Beton}}, \bibinfo
  {author} {\bibfnamefont {J.~P.}\ \bibnamefont {Oxley}}, \bibinfo {author}
  {\bibfnamefont {S.~P.}\ \bibnamefont {Beaumont}}, \bibinfo {author}
  {\bibfnamefont {S.}~\bibnamefont {Thoms}}, \ and\ \bibinfo {author}
  {\bibfnamefont {C.~D.~W.}\ \bibnamefont {Wilkinson}},\ }\href@noop {}
  {\bibfield  {journal} {\bibinfo  {journal} {Phys. Rev. Lett.}\ }\textbf
  {\bibinfo {volume} {64}},\ \bibinfo {pages} {2058} (\bibinfo {year}
  {1990})}\BibitemShut {NoStop}%
\bibitem [{\citenamefont {Sampietro}, \citenamefont {Fasoli},\ and\
  \citenamefont {Ferrari}(1999)}]{samp99}%
  \BibitemOpen
  \bibfield  {author} {\bibinfo {author} {\bibfnamefont {M.}~\bibnamefont
  {Sampietro}}, \bibinfo {author} {\bibfnamefont {L.}~\bibnamefont {Fasoli}}, \
  and\ \bibinfo {author} {\bibfnamefont {G.}~\bibnamefont {Ferrari}},\
  }\href@noop {} {\bibfield  {journal} {\bibinfo  {journal} {Rev. Sci.
  Instrum.}\ }\textbf {\bibinfo {volume} {70}},\ \bibinfo {pages} {2520}
  (\bibinfo {year} {1999})}\BibitemShut {NoStop}%
\bibitem [{\citenamefont {Molenkamp}\ and\ \citenamefont
  {de~Jong}(1994)}]{mole94}%
  \BibitemOpen
  \bibfield  {author} {\bibinfo {author} {\bibfnamefont {L.~W.}\ \bibnamefont
  {Molenkamp}}\ and\ \bibinfo {author} {\bibfnamefont {M.~J.~M.}\ \bibnamefont
  {de~Jong}},\ }\href@noop {} {\bibfield  {journal} {\bibinfo  {journal} {Phys.
  Rev. B}\ }\textbf {\bibinfo {volume} {49}},\ \bibinfo {pages} {5038}
  (\bibinfo {year} {1994})}\BibitemShut {NoStop}%
\bibitem [{\citenamefont {Maximov}\ \emph {et~al.}(2004)\citenamefont
  {Maximov}, \citenamefont {Gbordzoe}, \citenamefont {Buhmann}, \citenamefont
  {Molenkamp},\ and\ \citenamefont {Reuter}}]{maxi04}%
  \BibitemOpen
  \bibfield  {author} {\bibinfo {author} {\bibfnamefont {S.}~\bibnamefont
  {Maximov}}, \bibinfo {author} {\bibfnamefont {M.}~\bibnamefont {Gbordzoe}},
  \bibinfo {author} {\bibfnamefont {H.}~\bibnamefont {Buhmann}}, \bibinfo
  {author} {\bibfnamefont {L.~W.}\ \bibnamefont {Molenkamp}}, \ and\ \bibinfo
  {author} {\bibfnamefont {D.}~\bibnamefont {Reuter}},\ }\href@noop {}
  {\bibfield  {journal} {\bibinfo  {journal} {Phys. Rev. B}\ }\textbf {\bibinfo
  {volume} {70}},\ \bibinfo {pages} {121308(R)} (\bibinfo {year}
  {2004})}\BibitemShut {NoStop}%
\bibitem [{\citenamefont {Kristensen}\ \emph {et~al.}(1998)\citenamefont
  {Kristensen}, \citenamefont {Jensen}, \citenamefont {Zaffalon}, \citenamefont
  {Sorensen}, \citenamefont {Reimann},\ and\ \citenamefont
  {Lindelof}}]{kris98b}%
  \BibitemOpen
  \bibfield  {author} {\bibinfo {author} {\bibfnamefont {A.}~\bibnamefont
  {Kristensen}}, \bibinfo {author} {\bibfnamefont {J.~B.}\ \bibnamefont
  {Jensen}}, \bibinfo {author} {\bibfnamefont {M.}~\bibnamefont {Zaffalon}},
  \bibinfo {author} {\bibfnamefont {C.~B.}\ \bibnamefont {Sorensen}}, \bibinfo
  {author} {\bibfnamefont {S.~M.}\ \bibnamefont {Reimann}}, \ and\ \bibinfo
  {author} {\bibfnamefont {P.~E.}\ \bibnamefont {Lindelof}},\ }\href@noop {}
  {\bibfield  {journal} {\bibinfo  {journal} {J. Appl. Phys.}\ }\textbf
  {\bibinfo {volume} {83}},\ \bibinfo {pages} {607} (\bibinfo {year}
  {1998})}\BibitemShut {NoStop}%
\bibitem [{\citenamefont {Apetrii}\ \emph {et~al.}(2002)\citenamefont
  {Apetrii}, \citenamefont {Fischer}, \citenamefont {Kunze}, \citenamefont
  {Reuter},\ and\ \citenamefont {Wieck}}]{apet02}%
  \BibitemOpen
  \bibfield  {author} {\bibinfo {author} {\bibfnamefont {G.}~\bibnamefont
  {Apetrii}}, \bibinfo {author} {\bibfnamefont {S.}~\bibnamefont {Fischer}},
  \bibinfo {author} {\bibfnamefont {U.}~\bibnamefont {Kunze}}, \bibinfo
  {author} {\bibfnamefont {D.}~\bibnamefont {Reuter}}, \ and\ \bibinfo {author}
  {\bibfnamefont {A.}~\bibnamefont {Wieck}},\ }\href@noop {} {\bibfield
  {journal} {\bibinfo  {journal} {Semicond. Sci. Technol.}\ }\textbf {\bibinfo
  {volume} {17}},\ \bibinfo {pages} {735} (\bibinfo {year} {2002})}\BibitemShut
  {NoStop}%
\bibitem [{\citenamefont {Bockelmann}\ and\ \citenamefont
  {Bastard}(1990)}]{bock90}%
  \BibitemOpen
  \bibfield  {author} {\bibinfo {author} {\bibfnamefont {U.}~\bibnamefont
  {Bockelmann}}\ and\ \bibinfo {author} {\bibfnamefont {G.}~\bibnamefont
  {Bastard}},\ }\href@noop {} {\bibfield  {journal} {\bibinfo  {journal} {Phys.
  Rev. B}\ }\textbf {\bibinfo {volume} {42}},\ \bibinfo {pages} {8947}
  (\bibinfo {year} {1990})}\BibitemShut {NoStop}%
\bibitem [{\citenamefont {Shik}\ and\ \citenamefont {Challis}(1993)}]{shik93}%
  \BibitemOpen
  \bibfield  {author} {\bibinfo {author} {\bibfnamefont {A.~Y.}\ \bibnamefont
  {Shik}}\ and\ \bibinfo {author} {\bibfnamefont {L.~J.}\ \bibnamefont
  {Challis}},\ }\href@noop {} {\bibfield  {journal} {\bibinfo  {journal} {Phys.
  Rev. B}\ }\textbf {\bibinfo {volume} {47}},\ \bibinfo {pages} {2082}
  (\bibinfo {year} {1993})}\BibitemShut {NoStop}%
\bibitem [{\citenamefont {Das~Sarma}\ and\ \citenamefont
  {Campos}(1993)}]{das93}%
  \BibitemOpen
  \bibfield  {author} {\bibinfo {author} {\bibfnamefont {S.}~\bibnamefont
  {Das~Sarma}}\ and\ \bibinfo {author} {\bibfnamefont {V.~B.}\ \bibnamefont
  {Campos}},\ }\href@noop {} {\bibfield  {journal} {\bibinfo  {journal} {Phys.
  Rev. B}\ }\textbf {\bibinfo {volume} {47}},\ \bibinfo {pages} {3728}
  (\bibinfo {year} {1993})}\BibitemShut {NoStop}%
\bibitem [{\citenamefont {Kubakaddi}(2007)}]{kuba07}%
  \BibitemOpen
  \bibfield  {author} {\bibinfo {author} {\bibfnamefont {S.~S.}\ \bibnamefont
  {Kubakaddi}},\ }\href@noop {} {\bibfield  {journal} {\bibinfo  {journal}
  {Phys. Rev. B}\ }\textbf {\bibinfo {volume} {75}},\ \bibinfo {pages} {075309}
  (\bibinfo {year} {2007})}\BibitemShut {NoStop}%
\bibitem [{\citenamefont {Hansen}\ \emph {et~al.}(2001)\citenamefont {Hansen},
  \citenamefont {Kristensen}, \citenamefont {Pedersen}, \citenamefont
  {Soorensen},\ and\ \citenamefont {Lindelof}}]{hans01}%
  \BibitemOpen
  \bibfield  {author} {\bibinfo {author} {\bibfnamefont {A.~E.}\ \bibnamefont
  {Hansen}}, \bibinfo {author} {\bibfnamefont {A.}~\bibnamefont {Kristensen}},
  \bibinfo {author} {\bibfnamefont {S.}~\bibnamefont {Pedersen}}, \bibinfo
  {author} {\bibfnamefont {C.~B.}\ \bibnamefont {Soorensen}}, \ and\ \bibinfo
  {author} {\bibfnamefont {P.~E.}\ \bibnamefont {Lindelof}},\ }\href@noop {}
  {\bibfield  {journal} {\bibinfo  {journal} {Phys. Rev. B}\ }\textbf {\bibinfo
  {volume} {64}},\ \bibinfo {pages} {045327} (\bibinfo {year}
  {2001})}\BibitemShut {NoStop}%
\bibitem [{\citenamefont {Lin}\ \emph {et~al.}(2010)\citenamefont {Lin},
  \citenamefont {Lin}, \citenamefont {Chi}, \citenamefont {Chen}, \citenamefont
  {Ueda},\ and\ \citenamefont {Komiyama}}]{lin10}%
  \BibitemOpen
  \bibfield  {author} {\bibinfo {author} {\bibfnamefont {K.-T.}\ \bibnamefont
  {Lin}}, \bibinfo {author} {\bibfnamefont {Y.}~\bibnamefont {Lin}}, \bibinfo
  {author} {\bibfnamefont {C.~C.}\ \bibnamefont {Chi}}, \bibinfo {author}
  {\bibfnamefont {J.~C.}\ \bibnamefont {Chen}}, \bibinfo {author}
  {\bibfnamefont {T.}~\bibnamefont {Ueda}}, \ and\ \bibinfo {author}
  {\bibfnamefont {S.}~\bibnamefont {Komiyama}},\ }\href@noop {} {\bibfield
  {journal} {\bibinfo  {journal} {Phys. Rev. B}\ }\textbf {\bibinfo {volume}
  {81}},\ \bibinfo {pages} {035312} (\bibinfo {year} {2010})}\BibitemShut
  {NoStop}%
\bibitem [{\citenamefont {Buchholz}\ \emph {et~al.}(2010)\citenamefont
  {Buchholz}, \citenamefont {Fischer}, \citenamefont {Kunze}, \citenamefont
  {Bell}, \citenamefont {Reuter},\ and\ \citenamefont {Wieck}}]{buch10}%
  \BibitemOpen
  \bibfield  {author} {\bibinfo {author} {\bibfnamefont {S.~S.}\ \bibnamefont
  {Buchholz}}, \bibinfo {author} {\bibfnamefont {S.~F.}\ \bibnamefont
  {Fischer}}, \bibinfo {author} {\bibfnamefont {U.}~\bibnamefont {Kunze}},
  \bibinfo {author} {\bibfnamefont {M.}~\bibnamefont {Bell}}, \bibinfo {author}
  {\bibfnamefont {D.}~\bibnamefont {Reuter}}, \ and\ \bibinfo {author}
  {\bibfnamefont {A.~D.}\ \bibnamefont {Wieck}},\ }\href@noop {} {\bibfield
  {journal} {\bibinfo  {journal} {Phys. Rev. B}\ }\textbf {\bibinfo {volume}
  {82}},\ \bibinfo {pages} {045432} (\bibinfo {year} {2010})}\BibitemShut
  {NoStop}%
\end{thebibliography}%
\bibliographystyle{aipnum4-1}

\end{document}